\documentclass{article}
\usepackage{color}
 \usepackage{amsbsy}
  \usepackage{bm}
 \usepackage{amsfonts}
\usepackage{amssymb}
\usepackage{amsmath}
\usepackage{graphics}
\usepackage{graphicx}
\begin{document}

\title{In medium properties  of $K^0$ and $\phi$ mesons  under an external magnetic field }
\author{R. M. Aguirre}

\date{
\it{Departamento de Fisica, Facultad de Ciencias Exactas}, \\\it{Universidad Nacional de La Plata,} \\
\it{and IFLP, UNLP-CONICET, C.C. 67 (1900) La Plata, Argentina.}}

%\date{}

\maketitle

\begin{abstract}
The one loop polarization insertions for the kaon and the $\phi$
mesons are evaluated in a hadronic medium as functions of the
field intensity and the baryonic density. For this purpose an
effective chiral model of the hadronic interaction is used,
supplemented with a phenomenological $K-\phi$ interaction. The
propagators of the charged particles include the full effect of
the coupling to the magnetic field, and the anomalous magnetic
moments are taken into account for the baryon fields.  The
effective masses as well as the decay widths are introduced and
examined as functions of matter density and the magnetic
intensity.

\noindent
\\
%PACS: 21.30.Fe, 03.70.+k, 11.10.Wx, 03.65.Pm, 13.40.-f

\end{abstract}

\newpage

\section{Introduction}

The interaction between matter and strong magnetic fields is a
subject of permanent research \cite{LAI,MIRANSKY}. Among the
different empirical manifestations studied, it deserves a special
mention the  strong radiation coming from the activity of certain
astrophysical objects. In fact, the presence of intense magnetic
fields has been deduced from the observational data of a class of
neutron stars which have been described in the magnetar model
\cite{DUNCAN,THOMPSON}. The sustained X-ray luminosity in the soft
(0.5-10 keV) or hard (50-200 keV) spectrum, as well as the
bursting activity of these objects have been attributed to the
dissipation and decay of very strong magnetic fields. Their
intensity has been estimated around $10^{15}$ G at the star
surface, but could reach higher values in the dense interior. The
origin of the magnetism, however, is still under debate
\cite{RUDERMAN,THOMPSON2}.

As neutron stars are mainly composed of hadronic matter, the
knowledge of the hadronic properties under a strong magnetic field
is desirable to understand the composition and evolution of
magnetars. A considerable effort has been made in the last years
to study related aspects such as bulk properties of hadronic
matter \cite{CHAKRABARTY, SUH,MUKHERJEE}, the quark structure of
mesons by using lattice QCD \cite{LatticeQCD}, effective hadronic
models for pions
\cite{ANDERSEN,COLUCCI,CHEOUN,ADHYA,AGUIRRE,MANDAL} and vector
mesons \cite{KAWAGUCHI,GHOSH,MALLIK}.

The possible existence of kaon condensates in neutron stars has
largely been debated \cite{BROWN, GAL,KIM,PONS}. Kaons have a
significative role in the strong interaction because they are the
lightest particles with open strangeness and they have a similar
role as the pions have in the manifestation of the chiral
symmetry. The $\phi$ meson, instead, has closed strangeness and
its coupling to protons and neutrons would be excluded by the OZI
rule. Therefore, its coupling to the kaon field is the main
channel of interaction with the hadronic medium in the low energy
regime. For this reason the $\phi$ meson is one of the preferred
probes of the formation of a quark gluon plasma phase in heavy ion
collisions. Since it interacts scarcely with other hadrons, the
$\phi$ meson does not suffer considerable rescattering in the late
hadronization stage, decoupling earlier from the surrounding
hadronic medium \cite{SHOR}. In recent years the dynamics of the
$\phi$ meson has been theoretically investigated in relation to
the strange content of the nucleons \cite{GUBLER}, alternative
couplings \cite{CABRERA}, effective mass \cite{COBOS}, and exotic
nuclei \cite{COBOS2}. However, only a few works were devoted to
the study of magnetic fields on the properties of $K$ and $\phi$,
as for instance the $K$-meson coupling model of \cite{DEY} and the
quark-meson coupling model of \cite{YUE}.

The aim of this work is to investigate the effect of a uniform
external field on the properties of the $K$ and $\phi$ mesons,
immersed in  a hadronic medium under conditions relevant for the
study of neutron stars. In our approach the magnetic field is
treated as a classical external field, therefore we neglect
electromagnetic quantum corrections. We examine a range of matter
densities where only protons and neutrons are the relevant
baryons.

For this purpose we choose a model of the hadron interaction based
on a non-linear realization of the SU(3) chiral symmetry
\cite{MAO,MISHRA0,MISHRA1,KUMAR,MISHRA,SCHRAMM}, where pions and
kaons are regarded as Goldstone bosons. It also includes the
lightest scalar and vector mesons, whose masses are generated
dynamically, and a dilaton field to take account of the QCD trace
anomaly. A kaon condensed phase is predicted to occur for
densities greater than five times the normal nuclear density
($n_0$) \cite{SCHRAMM}. In others models of the kaon interaction
\cite{DEY,YUE,GLENDENNING} the transition point is  lowered to
approximately $3 \, n_0$. In any case, the presence of a magnetic
field has been determined to be the cause
of an increase of this threshold density \cite{DEY,YUE}.\\
In addition, we include here a derivative coupling between kaons
and $\phi$ meson \cite{KLINGL}.

The polarization insertion for the mesons is evaluated at the one
loop level. The propagators for the charged particles includes the
Landau quantum levels and, in the case of the baryons, also the
anomalous magnetic moments. The kaon field couples directly to the
baryons, hence the kaon selfenergy depends on the baryonic
density. In contrast for a low density equilibrium state and
assuming the OZI rules, the $\phi$ meson interacts only with the
kaons and the corrections to its propagator do not depend on the
matter density. This situation will cease as hyperons or a kaon
condensate become stable.

 The effective mass of the mesons in the hadronic
medium is defined and analyzed at zero temperature for a wide
range of magnetic intensities $10^{15}$ G $\leq B \leq 10^{19}$ G
and baryonic densities below three times the saturation density of
nuclear matter. The equilibrium conditions correspond to neutron
star matter, i.e. in equilibrium against beta decay and with zero
electric charge density. Hence additional lepton fields are
included.

This work is organized as follows. In the next section the
hadronic model is presented and the corresponding one-loop
polarization insertions are evaluated. The results and discussion
are given in Sec. III, and the conclusions are shown in Sec. IV.

\section{In-medium meson polarization insertion}\label{Sec1}

The effective model for the hadronic interaction described in
\cite{MAO} contains massless fields, which obtain their masses
 through the interaction with the scalar boson fields
$\sigma,\, \zeta$ and the iso-triplet $\bm{\delta}$.  There is
also a polynomial symmetry breaking term, and a scalar dilaton
field.

In the mean field approximation (MFA) each of these scalar fields is
decomposed as the sum of a vacuum expectation value  plus a
fluctuation. The only exception is the dilaton, for which zero
fluctuation is assumed. These mean values are partially absorbed in
the definition of the masses and in part they become parameters of
the residual interaction.

The kaon field, instead, appears as a Goldstone boson. Within the
same order of the chiral expansion it obtains non zero mass and a
derivative meson-baryon coupling given by

\begin{eqnarray}
\mathcal{L}&=& \frac{1}{8 f_K^2}\big(-3 i J^\mu
\bar{K}\stackrel{\leftrightarrow}{\partial}_\mu K - J^\mu_a
\bar{K} \tau_a \stackrel{\leftrightarrow}{\partial}_\mu K +4 d_1 J
\, \partial^\mu\bar{K} \partial_\mu K + 4 d_2 J_p \,
\partial^\mu K^+ \partial_\mu K^- \nonumber
\\
&+&4 d_2 J_n \,\partial^\mu\bar{K}^0 \partial_\mu
K^0\big)+\frac{m_K^2}{2 f_K}\big[ (\sigma+\sqrt{2} \zeta) \bar{K}
K + \bar{K}\bm{\delta} \cdot \bm{\tau} K \big] \nonumber \\
&+&\frac{1}{f_K}\big[ (\sigma+\sqrt{2} \zeta) \partial^\mu\bar{K}
\partial_\mu K +
\partial^\mu\bar{K}\bm{\delta} \cdot \bm{\tau}\partial_\mu K\big]
 \label{Lagran}
\end{eqnarray}

where the symbol $K$ without an index stands for the duplet
$\left(K^+,K^0\right)^t$, and $J^\mu=\bar{\Psi} \gamma^\mu \Psi$,
$J^\mu_a=\bar{\Psi} \gamma^\mu \tau_a \Psi$, the bi-spinor
$\Psi=\left(\psi_p, \psi_n\right)$ contains proton and neutron
fields. Furthermore $J_p=\bar{\psi}_p \psi_p$, $J_n=\bar{\psi}_n
\psi_n$, $J=J_p+J_n$. It must be stressed that $\sigma, \,
\bm{\delta}$ and $\zeta$ are used here for the fluctuations of the
scalar fields.

We complete Eq. (\ref{Lagran}) with an interaction between the
vector meson $\phi^\mu$ and the kaons \cite{KLINGL}

\begin{eqnarray}
\mathcal{L}=\frac{g}{\sqrt{2}} \; \phi^\mu \bar{K}\partial_\mu K
\label{Lagran2}
\end{eqnarray}

At the one loop order the kaon propagator is modified by the
diagrams shown in Fig. 1. The diagram (a) is first order and
corresponds to the baryonic tadpoles. After regularization it gives
zero vacuum contribution, and a finite density dependent part. The
diagram (b) corresponds to kaon tadpoles mediated by the scalar
mesons. Again its vacuum part is zero when it is properly
regularized, and it will contribute only in the case of kaon
condensation. Finally, diagram (c) corresponds to the one-meson
exchange . In the present approach only contributions coming from
mesons with masses below 1 GeV are considered and we focus on
situations were kaon condensation is not probable. Under these
conditions the one loop kaon polarization can be decomposed as a sum
of contributions coming from the different diagrams,
\begin{eqnarray}
\Pi_{K
\alpha}(p)&=&\Pi_\alpha^{(a)}(p)+\Pi_\alpha^{(b)}(p)+\Pi_\alpha^{(c)}(p)
\nonumber
\end{eqnarray}
\begin{eqnarray}
 \Pi^{(a)}_\alpha(p)&=&\frac{i}{2 f_K^2} \int
\frac{d^4q}{(2 \pi)^4} \Bigg\{  d_2 \;
\textsf{p}^2\left[\delta_{\alpha 1} \text{Tr}\,
G^{(p)}(q)+\delta_{\alpha 2} \text{Tr}\, G^{(n)}(q)
\right]\nonumber
\\
&+&\sum_{j=n,p}\left[\frac{1}{2}(3+I_\alpha I_j) p_\mu \text{Tr}\,
\gamma^\mu G^{(j)}(q)+d_1 \; \textsf{p}^2 \text{Tr}\,
G^{(j)}(q)\right]\Bigg\}  \label{Pol1}
\end{eqnarray}
\begin{eqnarray}
\Pi^{(b)}_\alpha(p)&=&\frac{i}{4
f_K^2}\left(2\textsf{p}^2-m_K^2\right)\left[
\Delta_\sigma(0)+I_\alpha I_{\alpha '}
\Delta^{(3)}_\delta(0)\right] \nonumber\\
&\times& \sum_{\alpha'}\int \frac{d^4q}{(2 \pi)^4}
\left(2\textsf{q}^2-m_K^2\right) \Delta_K^{(\alpha ')}(q)
\label{Pol2}
\end{eqnarray}
\begin{eqnarray}
\Pi^{(c)}_\alpha(p)&=&\frac{i}{f_K^2}\int \frac{d^4q}{(2 \pi)^4}
\left(p^\mu q_\mu-\frac{m_K^2}{2}\right)^2\Bigg\{ \left[
\Delta^{(1)}_\delta(p-q)+\Delta^{(2)}_\delta(p-q)\right] \Delta_K^{(\bar{\alpha})}(q)\nonumber\\
 &+&\left[ \Delta_\sigma(p-q)+ \Delta^{(3)}_\delta(p-q)\right]
\Delta_K^{(\alpha)}(q)\Bigg\}\label{Pol3}
\end{eqnarray}
where $\alpha=1 (2)$ corresponds to the charged (neutral) kaon,
$\bar{\alpha}=3-\alpha$, $I_k=(-1)^{1+k}$,
  $\textsf{p}^2=p_\mu
p^\mu$, $G$ stands for the nucleon propagators and $\Delta$ for
the scalar mesons propagators. If needed, a superscript indicates
the isospin component.

At the same order of approximation, the interaction of Eq.
(\ref{Lagran2}) induces the diagram of Fig. 1d for the $\phi$
meson propagator. The corresponding polarization insertion is
given by
\begin{eqnarray}
\Pi^{\mu \nu}_\phi(p)&=&-\frac{i}{2} g^2 \sum_\alpha \int
\frac{d^4q}{(2 \pi)^4} \Delta_K^{(\alpha)}(q)
\Delta_K^{(\alpha)}(q-p) (2 q-p)^\mu (2 q-p)^\nu \label{Pol4}
\end{eqnarray}

Since particles are immersed in a uniform magnetic field, we
include magnetic effects at the level of the propagators.
Therefore, for the charged mesons we use
\begin{eqnarray}
\Delta(p)=2 e^{i \Phi }\sum_n (-1)^n e^{-p_\bot^2/q B} L_n(2
p_\bot^2/q B) \left[\frac{1}{p_0^2-\omega_n^2+i \varepsilon}+2 \pi
i \delta(p_0^2-\omega_n^2) n_B(p_0) \right] \nonumber
\end{eqnarray}
where the phase factor $\Phi=q B(x+x')(y'-y)/2$ embodies the gauge
fixing, $\omega_n=\sqrt{m^2+p_z^2+(2 n+1)q B}$, and $n_B$ is the
statistical occupation function for bosons.

For the nucleon propagators we have to deal with spin, hence an
index $s=\pm 1$ is introduced to indicate spin projections along the
direction of the uniform magnetic field, and  we consider the
effects of the anomalous magnetic moments $\kappa$. For the neutron
we have
\begin{eqnarray}
G^{(n)}(p)= \sum_s \Lambda_s\left[\frac{1}{p_0^2-E_s^2+i\epsilon}+
2 \pi\,i\,n_F(p_0)\,\delta(p_0^2-E_s^2)\right] \nonumber
%\label{GN1}
\end{eqnarray}
 where $E_s=\sqrt{p_z^2+(\Delta-s\,\kappa_n B)^2}$,
 $\Delta=\sqrt{m_n^2+p^2_x+p^2_y}$, and
\begin{eqnarray}
\Lambda_s=\frac{ s}{2 \Delta}i\; \gamma^1 \gamma^2\left[ \not \!
u+ i \gamma^1 \gamma^2 (s \Delta-\kappa_n B)\right] \left( \not \!
v+m_n+ i s \Delta \gamma^1 \gamma^2\right)\nonumber % \label{GN2}
\end{eqnarray}
with $u_\mu=(p_0,0,0,p_z)$, and $v_\mu=(0,p_x,p_y,0)$.

The proton propagator is given by
\begin{equation}
G^{(p)}(p)=e^{i \Phi }  e^{-p_\bot^2/q B} \sum_{l,s}
\Lambda_{ls}\,\left[\frac{1}{p_0^2-E_{l s}^2+i\epsilon}+2
\pi\,i\,n_F(p_0)\,\delta(p_0^2-E_{l s}^2)\right]
\end{equation}
where $E_{l s}=\sqrt{p_z^2+(\Delta_l-s\,\kappa_p B)^2}$, $
\Delta_l=\sqrt{m_p^2+2 l q B}$, $\Lambda_{0 s}=(1+s)\left( \not \!
u+m_p-\kappa_p\,B\right) \Pi^{(+)}$, and
\begin{eqnarray}
\Lambda_{ls}&=& (-1)^l\frac{\Delta_l+s m_p}{ \Delta_l}\Big\{( \not
\! u-\kappa_p B+s \Delta_l) \Pi^{(+)} L_l(2 p_\bot^2/q B)
\nonumber\\
&-& ( \not \! u+\kappa_p B-s \Delta_l) \Pi^{(-)} \frac{s
\Delta_l-m_p}{s \Delta_l+m_p} L_{l-1}(2 p_\bot^2/q B)\nonumber \\
&+& \left[ \not \! u+i \gamma_1 \gamma_2 (s \Delta_l-\kappa_p
B)\right] i \gamma^1 \gamma^2 \not \! v \,\frac{s \Delta_l-
m_p}{2\,p_\bot^2} \left[ L_l(2 p_\bot^2/q B)-L_{l-1}(2 p_\bot^2/q
B)\right]\Big\} \nonumber % \label{DefGP2}
\end{eqnarray}
with $\Pi^{(\pm)}=(1\pm i\gamma^1\gamma^2)/2$ and $L_l$ stands for
the Laguerre polynomial of order $l$.

The propagators of the nucleons are constructed in a quasi particle
scheme, where the mass $m_k=m_0-g_\sigma S- g_\delta I_k D$ and the
energy spectrum $E_k+g_\omega W+g_\rho R I_k$ are modified by the
in-medium meson mean field values $S=g_\sigma
(N_{sp}+N_{sn})/m_\sigma^2, \; D=g_\delta
(N_{sp}-N_{sn})/m_\delta^2, \; W=g_\omega (N_p+N_n)/m_\omega^2, \;
R=g_\rho (N_p-N_n)/m_\rho^2 $, and
\begin{eqnarray}
N_p&=&\frac{q B}{2 \pi^2}\sum_{l,s}  \int dp_z \left[n_F(E_{l s},\bar{\mu}_p)-n_F(-E_{l s},\bar{\mu}_p)\right]\nonumber \\
N_n&=&\sum_s \int \frac{d^3p}{(2 \pi)^3}
 \left[ n_F(E_s,\bar{\mu}_n)-n_F(-E_s,\bar{\mu}_n)\right]\nonumber \\
N_{sp}&=&\frac{q B}{2 \pi^2} \, m\, \sum_{l, s} \int dp_z
\frac{\Delta_l+s\,\kappa_p
B}{E_{l s}\,\Delta_l}[n_F(E_{l s},\bar{\mu}_p)+n_F(-E_{l s},\bar{\mu}_p)] \nonumber \\
N_{sn}&=&\sum_s\int \frac{d^3p}{(2 \pi)^3}
 \frac{\Delta+s\,\kappa_n B}{E_s\,\Delta}\left[n_F(E_s,\bar{\mu}_n)+n_F(-E_s,\bar{\mu}_n)\right]\nonumber
\end{eqnarray}
The two first equations relate the conserved baryon number with
the chemical potentials $\mu_k$ by means of the effective chemical
potentials $\bar{\mu}_k=\mu_k-g_\omega W - g_\rho I_k R$.

The propagators just shown were derived within a thermal field
theory  \cite{AGUIRRE,A&D2016}, more specifically within the real
time formalism of Thermo-Field Dynamics. Here only the (1,1)
component is exhibited because it suffices for the present
calculations at zero temperature. These results combine the gauge
invariance of the proper time method \cite{SCHWINGER} with the
momentum representation of \cite{CHODOS}, furthermore they include
the contributions of the anomalous magnetic moments.

To isolate the divergences we use dimensional regularization for
neutral particles, and zeta function regularization
\cite{ELIZALDE} for charged ones. Physical contributions are
obtained after a subtraction procedure. For the divergent function
$\mathcal{F}(s)$ where $s=\textsf{p}^2$ and the regularization
point $s_0$, an N-order subtraction consist in the subtraction to
this quantity of $N+1$ terms
$\mathcal{F}^{(k)}(s_0)\;(s-s_0)^k/k!, \;\; 0 \leq k \leq N$.

After regularization, we obtain for the kaon polarizations
\begin{eqnarray}
\Pi^{(a)}_1(p)&=&-\frac{1}{2 f_K^2}\left\{p_0\,(2 N_p+ N_n)+
\textsf{p}^2 \left[(d_1+d_2) N_{sp}+d_1 N_{sn} \right]\right\}
\nonumber \\
\Pi^{(a)}_2(p)&=&-\frac{1}{2 f_K^2}\left\{p_0\,(N_p+2 N_n)+
\textsf{p}^2 \left[d_1 N_{sp}+(d_1+d_2) N_{sn} \right]\right\}
\nonumber
\end{eqnarray}
As mentioned before, we neglect the presence of a kaon condensate,
hence we obtain $\Pi^{(b)}_\alpha=0$. The long expressions for the
regularized $\Pi^{(c)}_2$ are shown in the Appendix A.

The polarization insertion for the $\phi$ meson can be written as
a sum of contributions coming from the neutral and charged kaons
\begin{eqnarray}
\Pi^{\mu \nu}_\phi(p)&=&\sum_\alpha\left(\mathcal{A}^{(\alpha)}
\frac{p^\mu p^\nu}{\textsf{p}^2} +\mathcal{B}^{(\alpha)} \, g^{\mu
\nu}\right) \label{PhiPol}
\end{eqnarray}
Details of the functions $\mathcal{A}, \, \mathcal{B}$ and their
decomposition into longitudinal and transversal blocks are shown  in
the Appendix B.

A Dyson-Schwinger approach is used to define the effective masses of
kaons and $\phi$ mesons as poles of the corresponding corrected
propagators. Hence the effective kaon mass $m_K^*$ corresponds to
the solutions of the equation
\begin{equation}
p_0^2-m_K^2-\text{Re} \, \Pi_K(p_0,\bm{p}=0)=0
\end{equation}
in the unknown $p_0$. For the decay width we use
$\Gamma_K=\text{Im}\, \Pi_K(\textsf{p}^2=m_K^{* 2})/m_K$ . \\
The definition of the effective mass for the $\phi$ meson requires
some care. Using the inverse propagator of a free vector meson
$D_0^{-1}=-\left(\textsf{p}^2-m_\phi^2\right)g^{\mu \nu}+p^\mu
p^\nu$, the Dyson Schwinger equation becomes
\begin{equation}
D_\phi^{-1}=-\left(\textsf{p}^2-m_\phi^2\right)g^{\mu \nu}+p^\mu
p^\nu-\left(\mathcal{A} \frac{p^\mu p^\nu}{\textsf{p}^2}
+\mathcal{B} \, g^{\mu \nu}\right)
\end{equation}
where we have assumed a structure similar to that described in
Eq.(\ref{PhiPol}) for the polarization insertion . In consequence we
can write
\begin{equation}
D_\phi^{\mu
\nu}(p)=\frac{1}{\textsf{p}^2-m_\phi^2+\mathcal{B}}\left(-g^{\mu
\nu}+\frac{\textsf{p}^2-\mathcal{A}}{m_\phi^2-\mathcal{A}-\mathcal{B}}\,
\frac{p^\mu p^\nu}{\textsf{p}^2} \right) \label{PhiProp}
\end{equation}
Thus we adopt the lowest solution of the equation
\begin{equation}
p_0^2-m_\phi^2+\text{Re}\,\mathcal{B}(p_0,\bm{p}=0)=0
\label{PhiMass}
\end{equation}
as the definition of the effective mass $m_\phi^*$. Furthermore, the
decay width is taken as $\Gamma_\phi=\text{Im}\,
\mathcal{B}(\textsf{p}^2=m_\phi^{* 2})/m_\phi$.\\It can be argued in
favor of this definition that in the proximity of the solution
$\textsf{p}^2\simeq m_\phi^{* \; 2}$ the propagator of
Eq.(\ref{PhiProp})behaves as
\begin{equation}
D_\phi^{\mu \nu}(p)\simeq\frac{1}{\textsf{p}^2-m_\phi^{* \;
2}}\left(-g^{\mu \nu}+ \frac{p^\mu p^\nu}{m_\phi^{* \; 2}} \right)
\nonumber
\end{equation}
namely a quasiparticle picture emerges, where the mass is dressed by
the kaon interaction.\\
As explained in the Appendix B, the functions $\mathcal{A}, \,
\mathcal{B}$ are different for the longitudinal and transversal
sectors. In the following we focus on the longitudinal meson branch.

\section{Results and discussion}

In this section we analyze the effective meson masses for a
hadronic environment which can be found in situations of
astrophysical interest. We consider electrically neutral matter in
equilibrium against weak decay under an external magnetic field.
We examine a range of matter densities below $7.5\times 10^{14}$
g/cm$^3$ at zero temperature, hence protons and neutrons are the
main components of the conserved baryonic number. Furthermore we
consider magnetic intensities  $10^{15}\leq B\leq10^{19}$ G, which
cover the empirical data attributed to magnetars and approach to
the QCD scale.

In this calculations we neglect the isotopic differences in the
kaon properties, adopting $m_K=495.5$ MeV and for the $\phi-K$
coupling we use $g=6.55$ an average of the values adjusted in
\cite{KLINGL} to reproduce the decay widths $\phi\rightarrow
K^0\bar{K}^0$ and $\phi \rightarrow K^+K^-$. In addition we use
$f_K=122$ MeV, $m_\phi=1020$ MeV,  $d_1=2.56/m_K$, $d_2=0.73/m_K$
\cite{MISHRA0} and a set of parameters  which predicts vacuum
values for the nucleon and meson masses $m_\sigma=466.5$ MeV,
$m_\delta=899.4$ MeV, $m_\zeta=1024.5$ MeV, $m_\omega=782.5$ MeV,
$m_\rho=763.$ MeV, and the hadronic couplings $g_\sigma=10.567, \,
g_\delta=2.487, \, g_\zeta=-0.461, \, g_\omega=13.326, \,
g_\rho=5.488$.\\
 This parametrization guarantees the KN scattering lengths and the binding
properties of nuclear matter in the MFA, the  saturation density
$n_0=0.15$ fm$^{-3}$, binding energy $E_B=-15.3$ MeV, symmetry
energy $E_s=31.6$  MeV, and slope parameter $L=65.9$ MeV
\cite{MISHRA1}.

As a first step we evaluate the MFA at zero temperature, obtaining
the chemical potentials, the effective nucleon mass, and the spin
polarization of matter as functions of the magnetic intensity and
the baryonic density.  The results of the MFA are inserted in the
neutron and proton propagators, to evaluate the
$\phi$ and $K^0$ polarization insertions.\\
In Fig. 2 we show some bulk properties obtained in the MFA. The
energy per particle with the mass subtracted is displayed in Fig.
2a as function of the density. Here we neglected the constant
contribution of the electromagnetic field. The magnetic
interaction is responsible of the bound state in the low density
regime. For the extreme case
case $B=10^{19}$ it extends beyond $n/n_0=2$.\\
In Fig. 2b the relative spin polarization is shown as a function
of the density, separately for protons and neutrons. Since the
magnetic field discriminates the spin orientation, we define for a
baryon $b$ the relative spin polarization
$W_s=(n_{b+}-n_{b-})/n_b$ where $n_b$ is the total density, and
$n_{b+}$  ($n_{b-}$) is the density of states with their third
component spin oriented in the same(opposite) direction of the
magnetic field. It can be see that the proton component is more
sensitive to the magnetic orientation, furthermore the statistical
effect of increasing the particle number attempts against the
order imposed by the external field.

Next we examine the effective mass of the neutral kaon as a
function of the baryonic density. In Fig. 3 the results
corresponding to $B=10^{18}$ G are presented. There we compare the
complete and the MFA treatments, the last one is obtained by
including the term $\Pi^{(a)}$ and the propagation of the
fluctuations of the scalar meson fields $\sigma, \; \zeta$ and
$\delta$ are replaced by their in medium expectation values. So
the kaon effective mass $m_K^*$ verify the following equations
(see \cite{MISHRA0})
\begin{eqnarray}
m_{K0}^{*\, 2}&=&m_K^2\pm \frac{m_{K0}^*}{2 f_K^2}\left(N_p+2
N_n\right)-\frac{m_K^2}{2 f_K}\left(\sigma+\sqrt{2}
\zeta-\delta\right)-\nonumber \\
&&\frac{m_{K0}^{*\, 2}}{2 f_K^2}\left[-2 f_K \left(\sigma+\sqrt{2}
\zeta-\delta\right)+d_1 N_{sp} +(d_1+d_2) N_{sn}\right]  \label{MFAK0}\\
m_{Kc}^{*\, 2}&=& m_K^2\pm \frac{m_{Kc}^*}{2 f_K^2}\left(2 N_p+
N_n\right)-\frac{m_K^2}{2 f_K}\left(\sigma+\sqrt{2}
\zeta+\delta\right)-\nonumber \\
&&\frac{m_{Kc}^{*\, 2}}{2 f_K^2}\left[-2 f_K \left(\sigma+\sqrt{2}
\zeta+\delta\right)+(d_1+d_2) N_{sp} +d_2 N_{sn}\right]
\label{MFAKp}
\end{eqnarray}
Solutions of these equations give the effective masses for the
neutral and charged kaons. The upper (lower) sign corresponds to
the particle (antiparticle) case.\\
This approach is  equivalent to the
 $B=0$ calculations carried out for instance in Ref. \cite{MISHRA0}.
Both results are very similar, exhibiting a monotonous increasing
behavior with density. Only for densities $n/n_0>1$ the
differences become appreciable.

 As explained in the Appendix A,
our full results include quantum corrections coming from the
vacuum due to the coupling to a pair of neutral mesons
($\Pi^{(c)}_{nn}$) and also due to two charged mesons
($\Pi^{(c)}_{cc}$). In the first case, there is no direct
dependence on the magnetic intensity $B$, so we can distinguish
the relative importance of this two terms by examining how the
effective mass depends on $B$. That is the subject of Fig. 4 where
the dependence of the kaon mass on the magnetic intensity at fixed
baryonic densities $n=0.075, \, 0.15$ and $0.3$ fm$^{-3}$ is shown
in a logarithmic scale.  For a given density, the effective mass
is almost constant for low intensities until a characteristic
strength around $B=10^{18} G$, where $m^*_K$ starts a pronounced
decrease. This reduction becomes steeper as the density grows, it
rounds $4\%, 2\%$ and $1\%$ for $n/n_0=0.5, 1$ and 2 respectively.
Thus  the effective kaon mass in the nuclear medium experiences
two opposite effects, it increases under the influence of the
matter density while strong magnetic fields try to reduce it. The
fact that $m^*_K/m_K>1$ under the conditions of this study, shows
that density effects are dominant, while magnetic effects
are moderate and become more appreciable at higher densities. \\
The behavior found for the kaon mass contrast with the results of
Ref.\cite{YUE}, where an schematic model of the quark confinement is
used. In that case a monotonous decrease with the baryonic density
is obtained, reaching a $20 \%$ reduction at $n=0.45$ fm$^{-3}$ for
neutron star matter. This remarkable difference can be attributed,
according to the previous discussion, to the  direct coupling
between kaons and baryons.

For the sake of completeness the density, as well as the magnetic
intensity dependence of the MFA masses are shown in Figs.5a  and
 5b respectively. We illustrate the cases of the $K^0$ and $K^+$
 components. From Fig.5a it can be seen that at very low density both effective
 masses tend to the degenerate vacuum value $m_K$.
As the density increases so do the effective masses, but more
moderately for the charged kaon. It must be pointed out that in
contrast, the effective masses of the antiparticles $\bar{K}^0$
and $K^-$ are decreasing functions of the density.\\
On the other hand in Fig.5b it is shown that the mass parameter is
almost constant for magnetic intensities below $10^{18}$ G for
both kaon iso-components. But beyond this value it becomes
slightly decreasing (increasing) for the neutral (charged) kaon.\\
In summary, within the MFA the effective masses of the duplet
$(K^+,K^0)$ are greater than its vacuum value for a wide range of
densities and magnetic intensities.

 The decay width of the $K^0$ within this model is strictly zero
 because we have not included $K-\pi$ couplings. We have verified
 that at this order of approximation, mesons heavier than pions do
 not open alternative channels of decay.

Finally we consider the decay width $\Gamma_\phi$ of the $\phi$
meson. In the absence of a kaon condensate, $\Gamma_\phi$ does not
depend on the matter density, hence the neutral mesons vertex
contributes with a constant value of $1.92$ MeV. The charged kaon
contribution instead, has a significative dependence on the
applied magnetic field  as can be seen in Fig. 6. The increase of
the intensity causes a blocking of this channel with a threshold
value $B_t$, given by $e B_t=m_\phi^2/4-m_K^2$, that is $B_t
\simeq 2.45 \times 10^{18}$ G. For $B < 4\times 10^{17}$ G, the
decay width exhibits rapid oscillations around $1.92$ MeV, i.e.
the same value as the neutral contribution. For stronger fields
the periodicity as well as the amplitude of the oscillations
increase quickly with $B$ until the threshold is reached. For
intensities slightly below $B_t$ the decay width increases as far
as $15 \%$ with respect to the $B=0$ case. The oscillatory
behavior  is a typical effect of the occupancy of the Landau
levels by the virtual mesons.

Up to this point we have developed a strict one loop approach in
the Dyson-Schwinger scheme, where the kaon field has a constant
and degenerate mass. A common practice in the study of the $\phi$
meson is to introduce an effective kaon mass $m_K^*$ depending on
the matter density, which has been obtained from a separate
calculation \cite{KO}, or by treating kaons in a different way as
compared with the remaining mesons \cite{COBOS2, GLENDENNING}.
\\Assuming the validity of this procedure one can estimate the behavior of
$\Gamma_\phi$ if the effective kaon masses obtained in the MFA
(Eqs. (\ref{MFAK0}) and (\ref{MFAKp})) were used in Eqs.
(\ref{ImBnn}) and (\ref{ImBcc}).\\
It is easy to understand that $\Gamma_\phi$ will be reduced in a
scenario where the kaon mass increases with density. It must be
taken into account that after numerical evaluation we found that
the real part of the $\phi$ meson polarization is smaller than
$m_\phi^2$ by at most one order of magnitude. Hence the solution
of Eq.(\ref{PhiMass}) coincides practically with the vacuum $\phi$
meson mass. Now consider the situation where $m_K^*$ increase with
the density, as for instance for the results shown in Fig.5a. By
considering Eq.(\ref{ImBnn}) with $m_K$ replaced by $m_K^*$ two
conclusions arise. In first place the neutral kaon contribution
becomes zero as well $m_K^*>m_\phi/2$. In the case analyzed in
Fig.5a this condition is verified for densities approximately
above one half the normal nuclear density for all the range of
magnetic intensities considered here. In second place, the
magnitude of Im$\mathcal{B}$ decreases as
$\left[1-(2 m_K^*/m_\phi)^2\right]^{3/2}$. \\
Similar conclusions can be stated for the charged kaon
contribution to the $\phi$ meson polarization by examining
Eqs.(\ref{ImBcc}) and (\ref{ImAcc}). But in this case the general
effect is emphasized by growing the field intensity due to the
shift $m_K^*+e\, B(2j+1)$. Furthermore the threshold intensity
$B_t$  will be pushed to lower values.\\
Thus, under these hypotheses we expect that the effect of a
growing kaon mass is to drastically reduce the decay width
$\Gamma_\phi$.\\ In opposition and by the same arguments one
expect
that it will be enhanced if $m_K^*/m_K<1$.\\
In any case, our calculations for the dependence of the $\phi$
meson decay width should be reproduced in the low density regime
of those variable kaon mass treatments.

\section{Conclusions}

In this work an analysis of the effective mass and decay width of
the lightest neutral mesons with strange content in a hadronic
environment and in the presence of an external magnetic field has
been carried out. As the results could depend on the composition
of the hadronic medium, we have taken a situation of astrophysical
interest, hadronic matter electrically neutral, in equilibrium
against weak decay at zero temperature.  We have focused on a
range of magnetic fields $10^{15}$ G $\leq B \leq 10^{19}$ G, and
baryonic densities which can be found in certain magnetars.
However, the matter density is not so high to have a significative
population of hyperons.

 The calculations have been made within a
covariant hadronic model based on chiral considerations, where kaons
are regarded as Goldstone bosons coupled to other bosons, including
a dilaton field. Their vacuum expectation values provide mass to the
interacting fields and the fluctuations are included to account for
quantum corrections. For this purpose we have used propagators which
include the full effect of the external magnetic field, and the
anomalous magnetic moments in the case of the baryons. Furthermore a
phenomenological interaction between kaons and the $\phi$ meson is
considered.

We have included one-loop corrections to the meson propagators. To
extract meaningful results we have performed a regularization
procedure of the polarization of the mesons subject to an external
magnetic field. The finite one-loop approximation to the
propagators is used to discriminate different contributions to the
effective masses and decay widths in a Dyson-Schwinger approach.

The in medium variation of the  effective mass of the $K^0$ meson
is dominated by density effects, while the dependence on the
magnetic intensity is moderate. We have found a monotonous
increase with the baryonic density, reaching a  $10 \%$
enhancement at three times the normal nuclear density.  The decay
width of the neutral kaon involving scalar $\sigma$ and $\delta$
mesons is zero.

Within the present approach the in-medium properties of the $\phi$
meson do not depend on the matter density in the absence of a kaon
condensate. Its decay width receives a constant contributions from
the $\bar{K}^0K^0$ channel of approximately $1.9$ MeV, while the
pair of charged kaons provides a characteristic behavior. It
varies rapidly around the same value $1.9$ MeV for relatively
small intensities, and drops to zero at the threshold strength
$B_t\simeq 2 \times 10^{18}$ G. The decay channel for the $K^-K^+$
pair is blocked for $B> B_t$ due to the impossibility to
accommodate a pair of virtual kaons within the discrete Landau
levels. \\
The $\phi$ meson plays an essential role in the debate about
neutron star structure, and consequently about magnetars. It has
been well established that the rise of the hyperon population in
the star matter is the cause of the softening of the hadronic
equation of state for medium to high densities. This fact, in
turn, reduces the maximum mass which can be attained by an
isolated neutron star. Some theoretical estimations put this upper
limit below the observational evidence of two solar masses
observed for instance in the binary star PSR J1614-2230
\cite{DEMOREST}. Thus the principle of minimum energy in star
matter requires the presence of hyperons, but calculations based
on this fact seems to be in contradiction with the observational
data. This complex situation has been known as
the {\it hyperon puzzle} and has been analyzed in detail in recent years \cite{CENTELLES,VIDANA,OERTEL}.\\
A possible solution of this problem involves the $\phi$ meson,
since it can induce additional repulsion in the hyperon-hyperon
(YY) channel without modifying the low density behavior because of
the OZI rule \cite{LIM, MARQUES}. The uncertainty about
hyperon-meson couplings gives room to speculation, and it is a
common practice to use experimental data on hyper-nuclei to model
the YY interaction \cite{FORTIN,GAL}. Furthermore, this scheme
relies on the effectiveness of the $\phi$ propagation in a dense
medium. Our results show that the external magnetic field
influence the stability of this meson. It is extraordinarily
stable for high intensities ($B > B_t$) and approximately
preserves its zero field decay width for $B < 10^{17}$ G. In
between, the magnetic field gives rise to additional
instabilities. If a similar behavior were detected for the
remaining channels of the YY interaction, we can infer that there
is a window of magnetic intensities $10^{17}$ G $<B< 2\times
10^{18}$ G where the configuration of stable
homogeneous star matter could differ from the zero field configuration.\\
Our predictions for $\Gamma_\phi$ does not depend on the
composition of matter if meson condensates are absent. Hence, in
the scenario of non-central heavy ion collisions our results
verifies that the $\phi$ meson is a faithful probe of the
hadronization process whenever the energy regime produces emergent
magnetic fields $B > B_t$, which is the range predicted for RHIC
experiments \cite{SKOKOV}. For field intensities within the range
expected for SPS experiences \cite{SKOKOV, MO} the effect of the
$K^-K^+$ decay channel is active but almost constant.

\section{Acknowledgements}
This work has been partially supported by a grant from the Consejo
Nacional de Investigaciones Cientificas y Tecnicas,  Argentina.

\section{Appendix A: $K^0$ polarization for the one-meson exchange}\label{AppA}

 The $K^0$ polarization contains vertices for either two neutral
mesons or two charged mesons
$\Pi^{(c)}_0=\Pi^{(c)}_{nn}+\Pi^{(c)}_{cc}$. In the first case the
extraction of physical outcome can be carried out by dimensional
regularization. Using standard procedures we obtain
\begin{eqnarray}
-32 \pi^2 f_K^2
\Pi^{(c)}_{nn}&=&\left(\frac{1}{\epsilon}-\frac{\gamma}{2}\right)\left(\textsf{p}^4-m_K^2
\textsf{p}^2+m_\chi^2 \textsf{p}^2
+m_K^4\right)+\frac{\textsf{p}^2}{6}\left(3m_K^2+3
m_\chi^2-\textsf{p}^2\right)\nonumber \\
&-&\int_0^1 dz \, \left[z(3 z-1)\textsf{p}^4+z m_\chi^2
\textsf{p}^2+(1-3 z)
m_K^2 \textsf{p}^2+\frac{1}{2}m_K^4 \right] \nonumber \\
&\times&\ln\left[-z(1-z)\textsf{p}^2+z m_\chi^2+(1-z)m_K^2\right]
+\mathcal{O}(\epsilon) \label{Reg0}
\end{eqnarray}
The structure of the divergence as $\epsilon \rightarrow 0$
suggests a second order subtraction procedure, from which we
obtain the regularized result
\begin{eqnarray}
\text{Im} \Pi^{(c)}_{nn}=\sqrt{(\textsf{p}^2+m_K^2-m_\chi^2)^2-4
m_K^2 \textsf{p}^2} \;
\frac{\left(\textsf{p}^2-m_\chi^2\right)^2}{64 \pi
f_K^2\textsf{p}^2} \Theta\left(\textsf{p}^2-(m_K+m_\chi)^2\right)
\nonumber
\end{eqnarray}
\begin{eqnarray}
64 \pi^2 f_K^2  \text{Re}
\Pi^{(c)}_{nn}=m_K^2(\textsf{p}^2-m_k^2)^2\int_0^1 dz
\,\left[\frac{z(z-1)}{m_\chi^2
z+m_K^2(1-z)^2}\right]^2\,\left[m_\chi^2
z+\left(\frac{3}{2}-4z+3z^2\right)m_K^2\right]
\nonumber \hspace{1.5cm} &&\\
+2 \int_0^1 dz  \left[z(3 z-1)\textsf{p}^4+z m_\chi^2
\textsf{p}^2+(1-3 z) m_K^2 \textsf{p}^2+\frac{1}{2}m_K^4 \right]
\, \ln\frac{|-z(1-z)\textsf{p}^2+z m_\chi^2+(1-z)m_K^2|}{m_\chi^2
z+m_K^2(1-z)^2}\hspace{0.5cm} &&\nonumber \\
-(\textsf{p}^2-m_k^2) \int_0^1 dz
\,z(z-1)\,\frac{m_K^4(1+2z-6z^2)+2 \textsf{p}^2
\left[(1-5z+6z^2)m_K^2+z m_\chi^2\right]}{m_\chi^2 z+m_K^2(1-z^2)}
\label{Reg2} \hspace{1.5cm} && \nonumber
\end{eqnarray}
 a sum over $\chi=\sigma, \, \delta $ is assumed, as stated in the second term between curly brackets in Eq.(\ref{Pol3}).

For $\Pi^{(c)}_{cc}$ we start with the first term between curly
brackets in Eq.(\ref{Pol3}),  it can be rewritten as
\begin{eqnarray}
\frac{4 i}{f_K^2} \sum_{n,l} \frac{(-1)^{n+l}}{(2 \pi)^4} \int
d^4q \left(p_\mu q^\mu-\frac{m_K^2}{2}\right)^2
e^{-q^2_\bot/\beta} L_n(2 q^2_\bot/\beta)e^{-k^2_\bot/\beta} L_l(2
k^2_\bot/\beta)\int_0^\infty dt\, t \int_0^1 dz e^{\Lambda t}
\label{AppA1}
\end{eqnarray}
where $\beta=q\,B, \; k=q-p$, $\Lambda=\textsf{u}_q^2-2 z
\textsf{u}_q \cdot \textsf{u}_p+z \textsf{u}_p^2-z
m_\delta^2-\beta-2 \beta z n-(1-z)(2
\beta l+m_K^2)$.\\%, $\textsf{v}_q^2=-q_\bot^2$, $m_c$ and $m_0$ are
%the masses of the charged and neutral meson respectively.
By changing the longitudinal variables of integration to
$\tilde{u}=u_q-z u_p$ followed by a Wick rotation, we obtain from
the exponential in Eq. (\ref{AppA1}) a gaussian factor. Furthermore,
the discrete index $n$ can be summed up by using the identity (see
Eq.(8.975.1) of Ref. \cite{G&R})
\begin{eqnarray}
\sum_{n=0}^\infty L_n(x) r^n =\frac{\exp [x\, r/(r-1)]}{1-r}
\nonumber
\end{eqnarray}
Thus Eq.(\ref{AppA1}) can be rewritten as
\begin{eqnarray}
&&\frac{-1}{f_K^2(2 \pi)^4} \int d^2\tilde{u}_E \int d^2q_\bot
\int_0^1 dz \int_0^\infty dt \, t \, \left[\left(\tilde{u}_\mu
u_p^\mu \right)^2+\left(z \textsf{u}_p^2+v_{q \, \mu}
v_p^\mu-\frac{1}{2}m_K^2 \right)^2\right]\,
e^{-t \tilde{u}^2_E}\nonumber \\
& \times& \exp\left\{t\left[z(1-z)\textsf{u}_p^2-z m_\delta^2-\beta
-(1-z)m_K^2\right]-\frac{q_\bot^2}{\beta} \tanh (\beta z
t)-\frac{k_\bot^2}{\beta} \tanh [\beta (1-z) t]\right\} \nonumber \\
& \times&\left[1+\tanh (\beta z t)\right]\left[1+\tanh \left(\beta
(1-z) t\right)\right] \nonumber
\end{eqnarray}
Since linear terms in $\tilde{u}$ within the square bracket of the
equation above will give zero contribution by symmetric integration,
they have been omitted.\\
 A change on the perpendicular variables $\tilde{v}=v_q-\alpha
 \, v_p$ with $\alpha^{-1}=1+\tanh (\beta z
t)\coth [\beta (1-z) t]$ leads to a product of a gaussian times a
polynomial in the new variables. Again linear terms will give zero
contribution after integration. Hence we obtain
\begin{eqnarray}
&&\frac{-1}{f_K^2(2 \pi)^4} \int d^2\tilde{u}_E \int d^2q_\bot
\int_0^1 dz \int_0^\infty dt \, t \, \left[\left(\tilde{u}_\mu
u_p^\mu \right)^2+\left(z \textsf{u}_p^2+ \alpha
\textsf{v}_p^2-\frac{1}{2}m_K^2\right)^2+\left(\tilde{v}_\mu v_p^\mu
\right)^2
\right]\nonumber \\
& \times& \exp\left\{t\left[z(1-z)\textsf{u}_p^2-z m_\delta^2
-(1-z)m_K^2-\beta \right]+ \alpha \textsf{v}_p^2 \tanh (\beta z
t)/\beta \right\}\,e^{\tilde{\textsf{v}}^2 \tanh [\beta (1-z)
t]/\alpha \beta} \, e^{-t \tilde{u}^2_E} \nonumber \\
&\times&\left[1+\tanh (\beta z t)\right]\left[1+\tanh \left(\beta
(1-z) t\right)\right] \nonumber
\end{eqnarray}
By performing the integrations in $\tilde{u}_E, \, \tilde{v}$ the
following expression is obtained
\begin{eqnarray}
&&\frac{ \beta}{32 \pi^2 f_K^2} \int_0^1 dz \int_0^\infty \frac{dt\,
\alpha}{\tanh [\beta (1-z) t]}
\left[\frac{\textsf{u}_p^2}{t}+\frac{\alpha \beta
\textsf{v}_p^2}{\tanh [\beta (1-z) t]}- 2 \left(z \textsf{u}_p^2+
\alpha  \textsf{v}_p^2-\frac{1}{2}m_K^2\right)^2
\right]\nonumber \\
& \times& \exp\left\{t\left[z(1-z)\textsf{u}_p^2-z m_\delta^2
-(1-z)m_K^2-\beta \right]+ \alpha \textsf{v}_p^2\tanh (\beta z
t)/\beta \right\} \nonumber \\
& \times&\left[1+\tanh (\beta z t)\right]\left[1+\tanh \left(\beta
(1-z) t\right)\right] \nonumber
\end{eqnarray}
 The integrand in this equation has  a singularity at $t=0$, to isolate the singularity
we use the same procedure as in the zeta function regularization
scheme \cite{ELIZALDE}. We introduce a regularization factor
$t^s$, with $0<s<1$ and anticipating $\textsf{v}_p^2=0$, a closed
form in terms of the Hurwitz zeta function $\zeta(z,q)$ can be
obtained for the first integration
\begin{eqnarray}
(2 \pi f_K)^2 \Pi^{(c)}_{cc}&=&-\frac{1}{4}\int_0^1 dz \left[
\left(z
\textsf{u}_p^2-\frac{1}{2}m_k^2\right)^2\Gamma(1+s)\zeta(1+s,\xi)-\beta\textsf{u}_p^2
\Gamma(s) \zeta(s,\xi)\right] \nonumber
\end{eqnarray}
where
\begin{eqnarray}
\xi&=&\frac{z m_\delta^2+(1-z)m_K^2+\beta-z(1-z)\textsf{u}_p^2 }{2
\beta} \nonumber
\end{eqnarray}

From this expression a single pole can be isolated as
$s\rightarrow 0$
\begin{eqnarray}
-(4 \pi f_K)^2 \Pi^{(c)}_{cc}&=&\int_0^1 dz \Bigg\{ \left[\beta
\textsf{u}_p^2 B_1(\xi)+\left(z
\textsf{u}_p^2-\frac{1}{2}m_k^2\right)^2 \right]\left(
\frac{1}{s}-\gamma\right)-\left(z
\textsf{u}_p^2-\frac{1}{2}m_k^2\right)^2\psi(\xi)
\nonumber \\
&&-\beta \textsf{u}_p^2 \ln\left[\frac{\Gamma(\xi)}{\sqrt{2
\pi}}\right]+\mathcal{O}(s)\Bigg\}
 \label{Reg1}
\end{eqnarray}
where $\psi, B_1$ and $\gamma$ are the di-gamma function, the
Bernoulli function and the Euler-Mascheroni constant respectively.
 It must be pointed out that
the coefficient of the pole is independent of $B$.\\
The function $\psi(\xi)$ is well defined for $\xi >0$, in fact
this is the case for the present calculations of the kaon
effective mass due to the expected range of $\textsf{u}_p^2$ .

Finally the following regularized version is obtained by making a
subtraction at $\textsf{u}_p^2=m_K^2$ and taking $s\rightarrow 0$

\begin{eqnarray}
 \Pi^{(c)}_{cc}&=&\frac{1}{(4 \pi f_K)^2}\int_0^1 dz
\Bigg\{\beta \textsf{u}_p^2
\ln\left[\frac{\Gamma(\xi)}{\Gamma(\xi_0)}\right] +\left(z
\textsf{u}_p^2-\frac{1}{2}m_k^2\right)^2\left[\psi(\xi)-\psi(\xi_0)\right]
\nonumber \\
&-&z(1-z)(1-2z)\left[4 z \textsf{u}_p^2-(2 z+1)m_K^2\right]
(\textsf{u}_p^2-m_K^2)\frac{m_K^2}{8 \beta} \psi'(\xi_0)
\nonumber\\&-&\frac{z^2 (1-z)^2}{32
\beta^2}\left(\textsf{u}_p^2-m_K^2\right)^2m_K^2\left[4 \beta
\psi'(\xi_0)+m_K^2(2 z-1)^2
\psi''(\xi_0)\right]\nonumber \\
&+&\frac{1}{2}z(1-z)\textsf{u}_p^2\left(\textsf{u}_p^2-m_K^2\right)\psi(\xi_0)\Bigg\}
\nonumber
\end{eqnarray}
where $ \xi_0=\left[z m_\delta^2+(1-z)^2m_K^2+\beta\right]/2\beta
$.

\section{Appendix B: One loop phi meson polarization}\label{AppB}

At the one loop level, the phi meson polarization includes a two
kaon vertex which can be either both neutral or both charged, so we
can write
$\Pi=\Pi_{nn}+\Pi_{cc}$.\\
The first term can be decomposed as
\begin{eqnarray}
\Pi_{nn}^{\mu \nu}&=&\frac{g^2}{32 \pi^2}\left[p^\mu p^\nu
\mathcal{A}(\textsf{p}^2)+g^{\mu \nu} \mathcal{B}(\textsf{p}^2)
\right] \label{AppB0}
\end{eqnarray}
the functions $\mathcal{A}, \,\mathcal{B}$ have single poles which
can be isolated by standard procedures of dimensional
regularization
\begin{eqnarray}
\mathcal{A}&=&\frac{1}{3}\left(\frac{2}{\epsilon}-\gamma
\right)-\int_0^1 dz \, \left(4 z^2-4 z+1\right)
\ln\left[-z(1-z)\textsf{p}^2+m_K^2\right]
+\mathcal{O}(\epsilon)\nonumber
\\
\mathcal{B}&=&-\frac{1}{3}\left(\frac{2}{\epsilon}+1-\gamma
\right)(\textsf{p}^2-6 m_K^2)+2 \int_0^1 dz \,
\left[z(1-z)\textsf{p}^2-m_K^2\right]
\ln\left[-z(1-z)\textsf{p}^2+m_K^2\right]
+\mathcal{O}(\epsilon)\nonumber
\end{eqnarray}
The meaningful results are obtained after subtraction of the
divergent terms at the regularization point
$\textsf{p}^2=m_\phi^2$, giving
\begin{eqnarray}
\mathcal{A}_{reg}&=& -\frac{i\pi}{3}\Theta\left( \textsf{p}^2 -4
m_K^2\right)\left(1-4\frac{m_K^2}{\textsf{p}^2}\right)^{3/2}-\int_0^1
dz \, \left(2 z-1\right)^2  \ln
\frac{z(1-z)\textsf{p}^2-m_K^2}{z(1-z)m_\phi^2-m_K^2}
%\\
%&+&\left( \textsf{p}^2-m_\phi^2\right)\int_0^1 dz
%\frac{z(1-z)(4z^2-4z-1)}{z(1-z)m_\phi^2-m_K^2}\nonumber
\\
\mathcal{B}_{reg}&=&\frac{i \pi}{3}\Theta\left(\textsf{p}^2-4
 m_K^2\right)\textsf{p}^2\left(1 -4 \frac{m_K^2}{\textsf{p}^2}\right)^{3/2}-\frac{1}{3}\left(\textsf{p}^2-m_\phi^2
\right)\nonumber
\\
&+&2 \int_0^1 dz \, \left[z(1-z)\textsf{p}^2-m_K^2\right]
\ln\frac{z(1-z)\textsf{p}^2-m_K^2}{z(1-z)m_\phi^2-m_K^2}
\label{ImBnn}
\end{eqnarray}

For $\Pi_{cc}$ we follow the same steps described  in the Appendix A
and arrive to the expression
\begin{eqnarray}
\Pi_{cc}^{\mu \nu}&=&\frac{2 g^2}{(2 \pi)^4} \int_0^1 dz
\int_0^\infty dt\, t\, \frac{\exp\left\{-2 \beta
t\xi+\frac{\alpha}{\beta}\tanh\left(\beta t z
\right)\textsf{v}_p^2\right\}}{(1-r)(1-e^{-2\beta t}/r) }\int d^2u_E
\; e^{-t u_E^2} \nonumber
\\
&\times&\int d^2v \exp\left\{\left[\tanh(\beta t
z)+\tanh\left(\beta t (1-z)\right)\right]
\textsf{v}^2/\beta\right\} I^{\mu \nu} \label{AppB1}
\end{eqnarray}
where $\xi=\left[m_K^2+\beta-z(1-z)\textsf{u}_p^2\right]/2\beta$,
$\alpha$ was defined in the Appendix A and
\begin{eqnarray}
I^{\mu \nu}&=&g^{\mu \nu}\left(-u_4^2 g^\mu_0+v_1^2 g^\mu_1+v_2^2
g^\mu_2+u_3^2  g^\mu_3\right)+(1-4 z+4 z^2)u_p^\mu u_p^\nu-\alpha (2
z-1) \left(u_p^\mu v_p^\nu+u_p^\nu v_p^\mu\right) \nonumber
\\
&\times&\left(1-\frac{\tanh(\beta t z)} {\tanh\left(\beta t
(1-z)\right)}\right)+\alpha^2 v_p^\mu v_p^\nu
\left(1-\frac{\tanh(\beta t z)}{\tanh\left(\beta t (1-z)\right)}
\right)^2
\end{eqnarray}
those terms giving zero contribution to either the integration
over $u_E$ or over $v$ have been omitted in this equation. An
analysis of Eq.(\ref{AppB1}) shows that, excepting the factor
$I^{\mu \nu}$, the integrand is invariant under the change of
variable $z \rightarrow 1-z$. In contrast $I^{\mu \nu}=(2
z-1)\lambda p^\mu p^\nu (1-\tanh\left(\beta t
(1-z)\right)/\tanh(\beta t z) )$ for $\mu=0,3$ and $\nu=1,2$, is
odd under the same change. Hence, integration over $z$ gives
$\Pi_{cc}^{\mu \nu}=0$ for the mixing
sector $\mu=0,3$ and $\nu=1,2$.\\
From the structure of $I^{\mu \nu}$ a decomposition similar to
that shown in Eq.(\ref{AppB0}) can be made for each of the
longitudinal or transversal components of $\Pi_{cc}^{\mu \nu}$. By
taking $\textsf{v}_p^2=0$, closed forms for $\mathcal{A}$ and
$\mathcal{B}$ in terms of the digamma function can be given. For
instance, in the longitudinal sector we have
\begin{eqnarray}
\mathcal{A}(\textsf{v}_p^2=0)=2 \beta \int_0^1 dz \; (1-2 z)^2
\int dt \frac{e^{-2 \beta t\xi}}{1-e^{-2 \beta t}} \label{AppB3}
\end{eqnarray}
As in the cases previously discussed, we introduce  a regularizing
factor $t^s, \; 0< s < 1$, which allows the integration over $t$
in terms of the Hurwitz zeta function $\zeta(1+s,\xi)$, whenever
the condition $4(m_K^2+\beta)-\textsf{u}_p^2>0$ is satisfied.
Otherwise, the parameter $\xi$ could become negative as $z$
varies. In such a case the result can be extended  by using the
functional relation
\begin{eqnarray}
\zeta(s,\xi)=\zeta(s,N+\xi)+\sum_{j=0}^{N-1}(j+\xi)^{-s}
\label{HURWITZ}
\end{eqnarray}
where $N$ is the minimum entire number which satisfies
$m_K^2+(2N+1)\beta-\textsf{u}_p^2/4>0$.

 After this decomposition, we expand the result
around $s=0$ obtaining
\begin{eqnarray}
\mathcal{A}(\textsf{v}_p^2=0)=\frac{1}{3}\left(\frac{1}{s}-\gamma\right)-\int_0^1
dz \,(1-2 z)^2\;
\left[\psi(\xi+N)-\sum_{j=0}^{N-1}\frac{\Gamma(1+s)}{(j+\xi)^{1+s}}\right]+\mathcal{O}(s)
\nonumber
\end{eqnarray}
In this formula an Sokhotski-Plemelj decomposition is used in each
term of the finite sum and finally the meaningful result is
obtained by regularizing at $\textsf{p}^2=m_\phi^2$
\begin{eqnarray}
\text{Re} \mathcal{A}_{reg}(\textsf{v}_p^2=0)&=&-\int_0^1 dz
\,(1-2 z)^2\; \left[\psi(\xi+N)-\psi(\xi_0+N)\right] +2
\sum_{j=0}^{N-1}\frac{D_j}{\textsf{u}_p^2} \ln
\left(\frac{1-D_j}{1+D_j}\right)\nonumber \\
\text{Im} \mathcal{A}(\textsf{v}_p^2=0)&=&-4\pi \beta
\sum_{j=0}^{N-1}\frac{D_j}{\textsf{u}_p^2} \Theta\left(1-4
\frac{m_K^2+\beta (2 j+1)}{\textsf{u}_p^2} \right) \label{ImAcc}
\end{eqnarray}
where $\xi_0=\left[m_K^2+\beta-z(1-z)m_\phi^2\right]/2 \beta$ and $D_j=\sqrt{1-4[m_K^2+\beta (2 j+1)]/\textsf{u}_p^2}$.\\
Proceeding in a similar way we obtain for $\mathcal{B}$
 \begin{eqnarray}
\mathcal{B}(\textsf{v}_p^2=0)&=&-\left(\frac{1}{s}-\gamma\right)\left(
\frac{\textsf{u}_p^2-6 m_K^2}{3}+4 \beta N\right)
-4 \beta \int_0^1 dz \, \ln\left[ \frac{\Gamma(\xi+N)}{\sqrt{2\pi}}\right] \nonumber\\
&+&4 \beta \sum_{j=0}^{N-1}\ln(\xi+j)+\mathcal{O}(s)
\end{eqnarray}
 In this equation it can be seen that, in opposition
to all the other cases studied in both Appendixes, there is an
explicit
dependence of the main coefficient on the magnetic field.\\
From the above equation the regularized contributions are
extracted as
\begin{eqnarray}
\text{Re} \mathcal{B}_{reg}(\textsf{v}_p^2=0)&=&-4 \beta\int_0^1
dz \, \left[\ln
\frac{\Gamma(\xi+N)}{\Gamma(\xi_0+N)}+\left(\textsf{u}_p^2-m_\phi^2\right)\psi(\xi_0)
\frac{z(1- z)}{2 \beta}-\sum_{j=0}^{N-1} \ln\left(\frac{\xi+j}{\xi_0+j}\right)\right] \nonumber \\
\text{Im} \mathcal{B}(\textsf{v}_p^2=0)&=&-\textsf{u}_p^2\;
\text{Im} \mathcal{A}(\textsf{v}_p^2=0) \label{ImBcc}
\end{eqnarray}

The imaginary parts of the polarizations are finite, so they do
not need to be regularized.

\newpage
\begin{figure}
\includegraphics[width=0.8\textwidth]{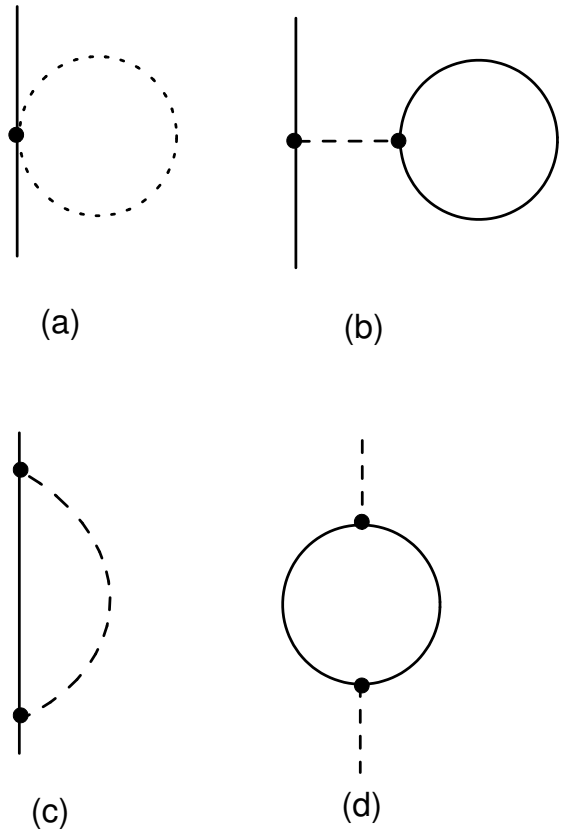}
\caption{\footnotesize Feynman graphs included in the present
calculations. Solid, dashed and dotted lines stand for kaon, non
strange scalar mesons and baryon propagators respectively. In the
case (d) the dashed line corresponds to the $\phi$ meson propagator.
}
\end{figure}
\newpage
\begin{figure}
\includegraphics[width=0.8\textwidth]{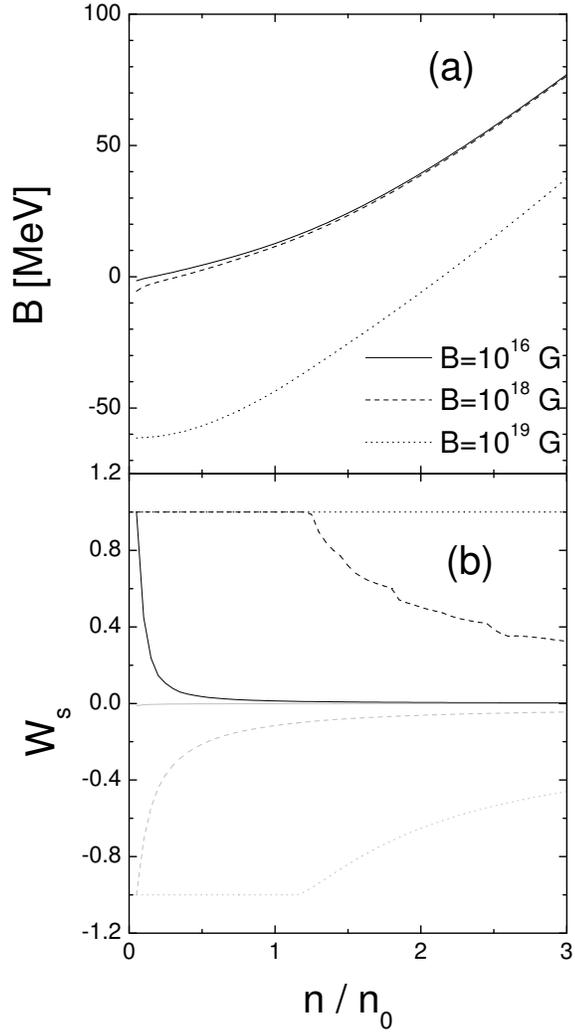}
\caption{\footnotesize Some characteristic bulk properties of
neutron star matter as a function of the density. The energy per
particle with the vacuum mass subtracted (a), and the spin
polarization for protons and neutrons (b) for several magnetic
intensities. In the last case dark (light) lines corresponds to
protons (neutrons).}
\end{figure}

\newpage
\begin{figure}
\includegraphics[width=0.8\textwidth]{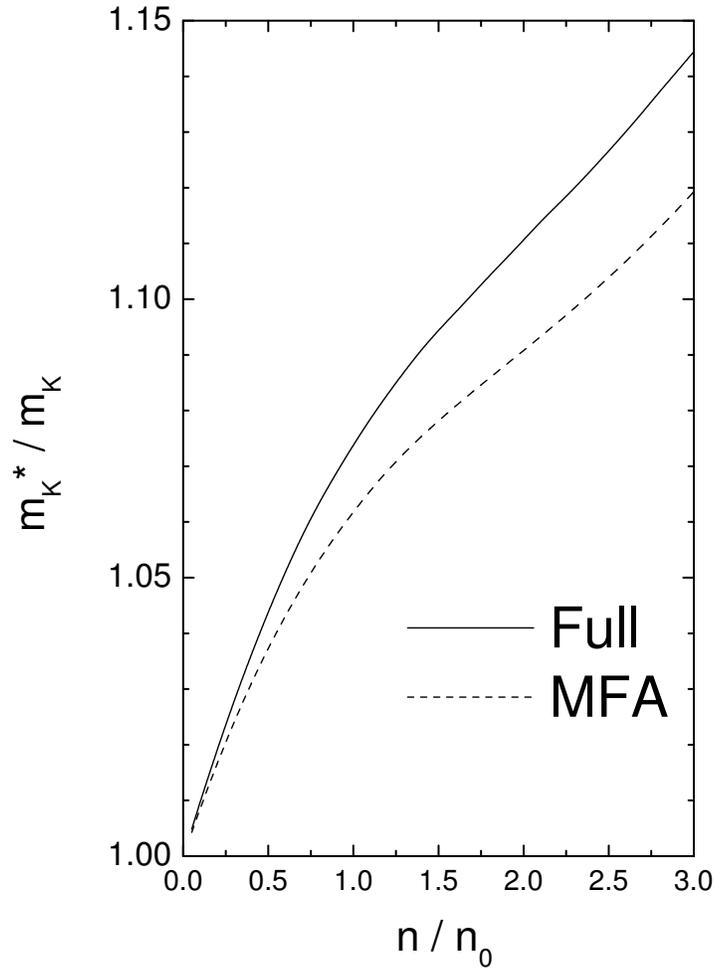}
\caption{\footnotesize The effective neutral kaon mass as a
function of the baryonic density at fixed magnetic intensity
$B=10^{18}$ G. Results corresponding to the mean field
approximation (MFA) and the full treatment are compared.}
\end{figure}

\newpage
\begin{figure}
\includegraphics[width=0.8\textwidth]{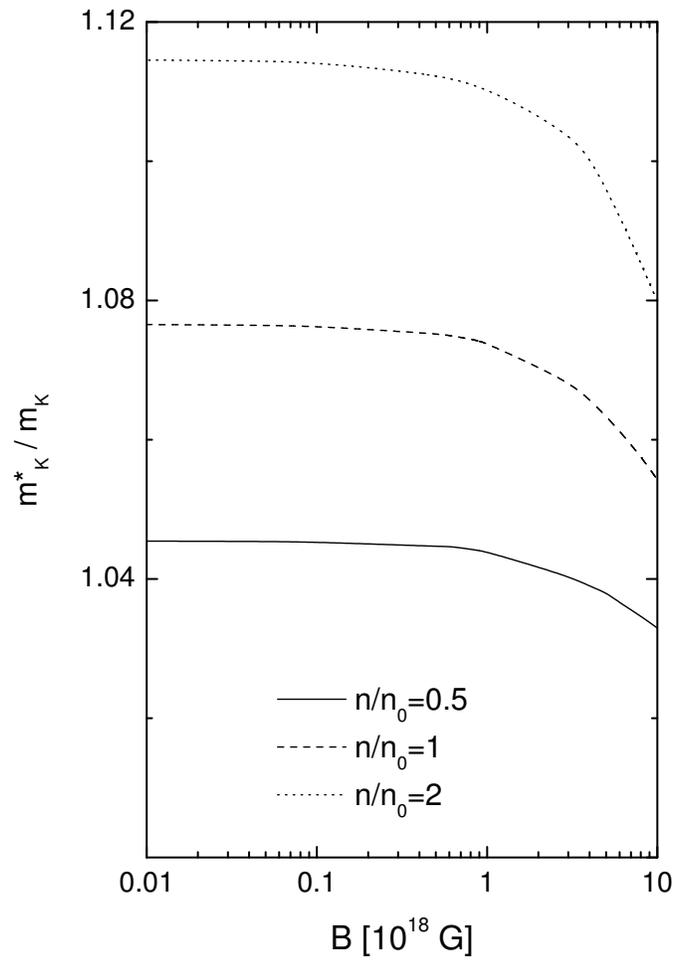}
\caption{\footnotesize The effective kaon mass as a function of
the magnetic intensity for the fixed baryonic densities
$n/n_0=0.5, 1,$ and $2$.}
\end{figure}

\newpage
\begin{figure}
\includegraphics[width=0.8\textwidth]{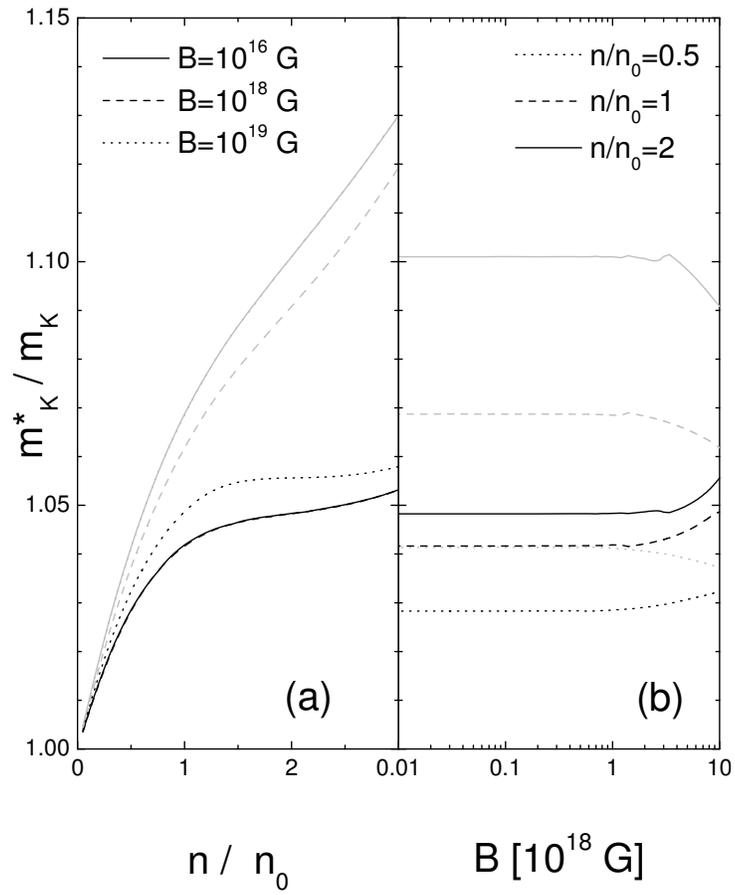}
\caption{\footnotesize The effective kaon masses in the MFA as a
function of the density (a) and its dependence on the magnetic
intensity (b). Dark (light) lines correspond to $K^+$ ($K^0$)
case.}
\end{figure}

\newpage
\begin{figure}
\includegraphics[width=0.8\textwidth]{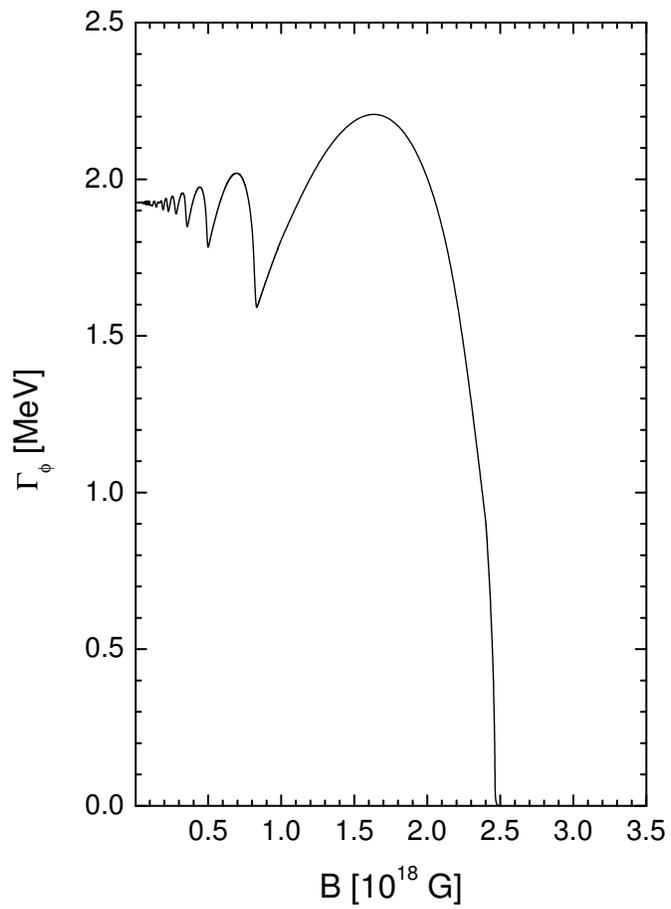}
\caption{\footnotesize Charged meson contribution to the $\phi$
meson decay width as a function of the magnetic intensity.}
\end{figure}

\end{document}